\documentclass[journal]{IEEEtran} 

\usepackage[cmex10,intlimits]{amsmath}
\usepackage{amsfonts}
\usepackage{color,colortbl}
\usepackage[pdftex]{graphicx}
\usepackage{tikz}

\usepackage{cite}
\usepackage{url}
\usepackage{hyperref}
\usepackage{multirow}
\usepackage{booktabs}

\usepackage{subfigure}
\usepackage{units}
\usepackage{etoolbox}
\usepackage{animate}

\providecommand*{\mrm}[1]{\mathrm{#1}}

\DeclareMathAccent{\ring}{\mathalpha}{operators}{"17}

\providecommand*{\unit}[1]{\ensuremath{\mrm{\,#1}}}

\providecommand*{\ju}{\ensuremath{\mrm{j}}}

\providecommand*{\diff}{\operatorname{d}\!}

\newcommand{\partder}[2]{\frac{\partial#1}{\partial#2}}

\newcommand{\totder}[2]{\frac{\diff #1}{\diff #2}}

\renewcommand{\vec}[1]{{\boldsymbol#1}}
\newcommand{\mat}[1]{{\mathbf{#1}}}
\newcommand{\R}{\mathbb{R}}

\newcommand{\eig}{\mathop{\mrm{eig}}}
\providecommand*{\diffS}{\operatorname{dS}\!}

\newcommand{\reg}{\varOmega}

\newcommand{\rv}{\vec{r}}
\newcommand{\psiv}{\vec{\psi}}

\newcommand{\Jm}{\mat{I}}

\newcommand{\Zm}{\mat{Z}}
\newcommand{\Xm}{\mat{X}}
\newcommand{\Rm}{\mat{R}}
\newcommand{\Om}{\mat{0}}

\newcommand{\Tm}{\mat{T}}

\newcommand{\We}{W_{\mrm{e}}}
\newcommand{\Wm}{W_{\mrm{m}}}

\newcommand{\Xma}{\mat{X}_{\alpha\nu}}
\newcommand{\Xme}{\mat{X}_{\mrm{e}}}
\newcommand{\Xmm}{\mat{X}_{\mrm{m}}}

\newcommand{\herm}{\mrm{H}}
\newcommand{\ellx}{\ell_\mrm{x}}
\newcommand{\elly}{\ell_\mrm{y}}
\newcommand{\Rml}{\mat{R}_{\Omega}}
\newcommand{\Rmr}{\mat{R}_{\mrm{r}}}

\providecommand*{\constraint}[1]{\ensuremath{\bar{#1}}}
\newcommand{\Pl}{P_{\mrm{\Omega}}}

\newcommand{\Prad}{P_{\mrm{r}}}
\newcommand{\Plc}{\constraint{P}_{\mrm{\Omega}}}
\newcommand{\Pwc}{\constraint{P}_{\mrm{w}}}

\newcommand{\ie}{\textit{i.e.}}
\newcommand{\eg}{\textit{e.g.}}
\newcommand{\cf}{\textit{cf.}}

\newcommand{\Rsurf}{R_\mrm{s}}

\newcommand{\minimize}{\mrm{minimize}}

\newcommand{\maximize}{\mrm{maximize}}

\newcommand{\subto}{\mrm{subject\ to}}

\newcommand{\eigv}{\gamma}
\newcommand{\Ohm}{\unit{\Omega}}

\newcommand{\Jv}{\vec{J}}
\newcommand{\Ov}{\vec{0}}
\newcommand{\Lm}{\mat{L}}
\newcommand{\Cmi}{\mat{C}_{\mrm{i}}}
\newcommand{\Rmre}{\mat{R}_{\mrm{TM}}}
\newcommand{\Rmrm}{\mat{R}_{\mrm{TE}}}

\newcommand{\gammame}{\boldsymbol{\gamma}_{\mrm{e}}}
\newcommand{\gammamm}{\boldsymbol{\gamma}_{\mrm{m}}}
\newcommand{\QL}{Q_\mrm{ub,L}} 
\newcommand{\Qrad}{Q^\mathrm{rad}}
\newcommand{\Qradlb}{\Qrad_\mathrm{lb}} 
\newcommand{\QradlbTM}{\Qrad_\mathrm{lb,TM}} 
\newcommand{\QradlbTE}{\Qrad_\mathrm{lb,TE}}
\newcommand{\QradTM}{\Qrad_\mathrm{TM}}
\newcommand{\QradTE}{\Qrad_\mathrm{TE}}

\hypersetup{
	colorlinks=true, 
	linkcolor=blue!70!black, 
	citecolor=red!70!black, 
	filecolor=blue!70!black, 
	urlcolor=blue!70!black, 
}

\begin{document}
\title{Trade-off Between Antenna Efficiency and Q-Factor}
\author{Mats~Gustafsson,~\IEEEmembership{Senior Member,~IEEE,}
        Miloslav~Capek,~\IEEEmembership{Senior Member,~IEEE,}	    
	    and Kurt~Schab,~\IEEEmembership{Member,~IEEE}
\thanks{Manuscript received  \today; revised \today.
This work was supported by the Swedish Foundation for Strategic Research (SSF) and by the Czech Science Foundation under Project 15-10280Y.  Kurt Schab is supported by the United States Intelligence Community Postdoctoral Research Fellowship Program.
}
\thanks{M.~Gustafsson is with the Department of Electrical and Information Technology,
Lund University, 221~00 Lund, Sweden (e-mail: mats.gustafsson@eit.lth.se).}
\thanks{M.~Capek is with the Department of Electromagnetic Field, Faculty of Electrical Engineering, Czech Technical University in Prague, 166~27 Prague, Czech Republic (e-mail: miloslav.capek@fel.cvut.cz).}
\thanks{K.~Schab is with the Department of Electrical and Computer
Engineering, Antennas and Electromagnetics Laboratory, North Carolina State University, Raleigh, NC, USA (e-mail: krschab@ncsu.edu).}
}

\maketitle

\begin{abstract}
The trade-off between radiation efficiency and antenna bandwidth, expressed in terms of Q-factor, for small antennas is formulated as a multi-objective optimization problem in current distributions of predefined support. Variants on the problem are constructed to demonstrate the consequences of requiring a self-resonant current as opposed to one tuned by an external reactance. The resulting Pareto-optimal sets reveal the relative cost of valuing low radiation Q-factor over high efficiency, the cost in efficiency to require a self-resonant current, the effects of lossy parasitic loading, and other insights.
\end{abstract}


\IEEEpeerreviewmaketitle

\section{Introduction}
\label{sec:introduction}

The radiation efficiency and bandwidth, the two of the most important antenna performance parameters~\cite{Volakis+etal2010}, are known to be strongly affected by the size of the radiator.  Because bandwidth can be increased arbitrarily through resistive loading, an antenna's bandwidth-efficiency product, $B\eta$, is a useful metric for assessing an antenna's bandwidth independent of loading since it is approximately inversely proportional to the antenna's radiation Q-factor, expressed here as the ratio of total Q-factor and efficiency, $Q/\eta$. Here we pose a question about the trade-off between these quantities: \textit{By looking for the most efficient antenna possible, do we sacrifice radiation Q-factor?}

A plethora of previous approaches exist for studying the fundamental bounds on antenna efficiency and bandwidth~\cite{Gustafsson+etal2015b}, rare, however, are attempts to investigate their mutual relationship~\cite{Gustafsson2013a}. Given the relation between Q-factor and bandwidth~\cite{Yaghjian+Best2005, Gustafsson+Nordebo2006, Capek+etal2015a}, a majority of the works treating the principal limits of antenna bandwidth deal with Q-factor~\cite{IEEEantennaterms1993}, as its useful single-frequency substitute. A key step in determining antenna Q-factor from single-frequency data is the evaluation of the system's stored energy~\cite{Schab+etal2018}. Pioneering work on this problem involved techniques based on equivalent circuits~\cite{Chu1948, Thal1978}, determination of electromagnetic fields generated by the radiator and its analytical integration~\cite{Collin+Rothschild1964, Hansen+Collin2009}. Practical and direct estimation of Q-factor from port quantities like input impedance~\cite{Kajfez+Wheless1986, Yaghjian+Best2005, Gustafsson+Jonsson2015a} also appeared, though these methods cannot be used to study bounds.  Recently, the stored energy of electrically small radiators was formulated as a functional of surface current density~\cite{Vandenbosch2010, Jonsson+Gustafsson2015}, by which means a missing link to its matrix-form definition has been traced back to Harrington,~\cite{Harrington+Mautz1972, Gustafsson+Nordebo2013}. Casting these energy and power functionals as quadratic forms under a method of moments approximation makes it possible to reformulate the optimization problem of finding the lower bound on Q-factor into its dual form~\cite{Gustafsson+etal2016a}, which is solvable by a parametrized eigenvalue problem~\cite{Capek+etal2017b}. Other methods to obtain Q-factor bounds include approaches based on forward scattering~\cite{Gustafsson+etal2007a}, electrically small limits~\cite{Yaghjian+Stuart2010,Vandenbosch2011,Yaghjian+etal2013}, and circuit models~\cite{Thal2012}, see~\cite{Gustafsson+etal2015b} for an overview. In general, these methods agree for electrically small structures where the significance and importance of Q-factor bounds are highest.

Research into the bounds on radiation efficiency is notably less prolific. Albeit the general understanding came already from the classical works~\cite{Hansen1981, Volakis+etal2010},  rigorous quantitative studies focus only on spherical geometries~\cite{Arbabi+SafaviNaeini2012, Karlsson2013, Fujita+Shirai2015, Fujita+Shirai2017}. Together with estimation of the bounds, there are practically-oriented contributions~\cite{Sarrazin+etal2016, Karlsson+Carlsson2009}, attempting to produce realistic antennas with maximal efficiency. A modal decomposition was used to estimate the upper limit on efficiency~\cite{Capek+etal2015b}. It was followed by practically-oriented heuristic optimization and discussion~\cite{Li+etal2017a, Li+etal2018a}. Other estimations are based on assumption of constant current~\cite{Shahpari+Thiel2016}, which is believed to maximize the efficiency, or utilization of circuit theory \cite{Pfeiffer2017,Thal2018}. Recently, the lower bounds of efficiency of self-resonant current were found using Lagrange multipliers~\cite{Jelinek+Capek2017} in a similar manner as in~\cite{Uzsoky+Solymar1956, Harrington1965} for externally tuned currents. In contrast to Q-factor, radiation efficiency is negatively influenced by utilization of the matching circuit~\cite{Smith1977, Jelinek+etal2017}, which should be therefore taken into account.

Here, we recast the optimization problems used to determine fundamental bounds on antenna Q-factor~\cite{Capek+etal2017b} and efficiency~\cite{Harrington1965} into multi-objective form. By doing so, we observe the cost incurred in each parameter under different objective weights, \ie{}, Pareto optimality~\cite{Boyd+Vandenberghe2004}. We also study the behavior of this cost under practical constraints previously shown to impact the fundamental bounds on efficiency or Q-factor individually.  These constraints include a requirement of self-resonance~\cite{Smith1977, Jelinek+etal2017} and the inclusion of non-controllable parasitic bodies~\cite{Cismasu+Gustafsson2014a}. Further, we analyze the frequency dependence of the trade-off between efficiency and Q-factor, both in terms of limiting cases and the complete Pareto front.

The optimization task itself relies on several crucial steps, \eg{}, expressing all parameters as quadratic forms of underlying matrix operators and determining a region in which all the operators are positive semidefinite~\cite{Gustafsson+etal2016a}. However all these tasks can be executed quickly and, importantly, the entire process is strictly deterministic (no heuristic algorithms are needed). In all cases considered in this paper, the optimization problems are transformed into convex form, implying their solution is global and efficiently found using convex optimization algorithms \cite{Boyd+Vandenberghe2004}.  In fact, many of the optimization problems discussed in this paper are solvable in closed form via eigenvalue problems. The main outputs of these routines are not only the bounds themselves, but also the optimal current densities associated with the bounds. Thus the results from this paper provide quantitative and physical insight into the question posed above, as well as its complement. 

The rest of the paper is organized as follows: the method-of-moments forms required to calculate the efficiency and Q-factors of arbitrary current distributions are introduced in Section~\ref{sec:background}. The multi-objective optimization problems are expressed and solved in Section~\ref{sec:pareto}.  The properties of the Pareto fronts are discussion in Section~\ref{sec:degen}, and compared with already known limiting cases in Section~\ref{sec:comparison}. Limiting cases in the low-frequency limit are discussed in Sec.~\ref{sec:sizelimits}. A practical example in which only part of the structure is controllable by the designer is shown in Section~\ref{sec:subregion}. The paper is concluded in Section~\ref{sec:conclusions}.

\section{Quadratic Forms For Optimized Quantities}
\label{sec:background}

A standard method-of-moments~(MoM) implementation of the electric field integral equation (EFIE) is used to compute the impedance matrix of an object~\cite{Harrington1968}
\begin{equation}
\Zm=\Rm+\ju\Xm=\Rmr+\Rml+\ju(\Xmm-\Xme).
\label{eq:Zmat}
\end{equation}
Here, vector functions on the object under consideration are approximated and expanded using a suitable set of basis functions $\{\psiv_m(\rv)\}$. The time-harmonic domain with~$\omega$ as the angular frequency is used throughout the paper. The MoM approximation of the stored magnetic and electric energies are~\cite{Vandenbosch2010,Harrington+Mautz1972,Gustafsson+etal2016a}
\begin{equation}
\Wm \approx\frac{1}{8}\Jm^{\herm}\left(\partder{\Xm}{\omega}+\frac{\Xm}{\omega}\right)\Jm = \frac{1}{4\omega}\Jm^{\herm}\Xmm\Jm 	
\label{eq:WeJV}
\end{equation}
and
\begin{equation}
\We \approx \frac{1}{8}\Jm^{\herm}\left(\partder{\Xm}{\omega}-\frac{\Xm}{\omega}\right)\Jm = \frac{1}{4\omega}\Jm^{\herm}\Xme\Jm, 
\label{eq:WmJV}
\end{equation}
respectively, where $\Jm$ denotes the column matrix of MoM expansion coefficients~\cite{Harrington1968} and $^\herm$ denotes Hermitian conjugate. The Q-factor of the current represented by $\Jm$ reads~\cite{IEEE145-1993}
\begin{equation}
Q = \frac{2\omega\max\left\{\Wm,\We\right\}}{\Prad+\Pl} \approx \eta\frac{\max\left\{\Jm^{\herm}\Xmm\Jm,\Jm^{\herm}\Xme\Jm\right\}}{\Jm^{\herm}\Rmr\Jm},
\label{eq:Q}
\end{equation}
where $\eta$ denotes the radiation efficiency 
\begin{equation}
\eta = \frac{\Prad}{\Prad+\Pl} =\frac{1}{1+\delta} \approx \frac{\Jm^{\herm}\Rmr\Jm}{\Jm^{\herm}(\Rmr+\Rml)\Jm},
\label{eq:eff}
\end{equation}
$\Prad$ the radiated power, $\Pl$ the power dissipated from ohmic losses, and $\delta=\Pl/\Prad$ the dissipation factor~\cite{Harrington1960}. The radiation Q-factor $\Qrad=Q/\eta$ represents the Q-factor of the current independent of losses.

The dissipated power due to these ohmic losses on an object with surface resistance $\Rsurf(\rv)$ is 
\begin{equation}
\Pl\approx\frac{1}{2}\Jm^{\herm}\Rml\Jm
\label{eq:losses-nonuniform}
\end{equation}
where
\begin{equation}
\mathbf{R}_{\Omega,mn} = \int_{\reg}\psiv_m(\rv)\cdot\psiv_n(\rv)\Rsurf(\rv)\diffS.
\label{eq:lossymatrix}
\end{equation}
In the special case when the surface resistance is uniform over the entire object, \ie{}, $\Rsurf(\rv) = \Rsurf$, calculation of the dissipated power reduces to
\begin{equation}
\Pl\approx\frac{\Rsurf}{2}\Jm^{\herm}\boldsymbol{\Psi} \Jm,
\label{eq:losses}
\end{equation}
where $\boldsymbol{\Psi}$ is the Gram matrix of the basis-functions
\begin{equation}
\Psi_{mn} = \int_{\reg}\psiv_m(\rv)\cdot\psiv_n(\rv)\diffS.
\label{eq:overlapmatrix}
\end{equation}
The resistance $\Rsurf$ is commonly modeled with resistive sheets and surface resistance as~\cite{Senior+Volakis1995}
\begin{equation}
\Rsurf = \frac{1}{\sigma d}\quad\mathrm{and}\quad \Rsurf = \sqrt{\frac{\omega\mu_0}{2\sigma}},
\label{eq:sheet} 
\end{equation}
respectively, where~$\sigma$ is the conductivity,~$d$ sheet thickness, and~$\mu_0$ free space permeability. The corresponding reactance in the skin depth model can be incorporated in the reactance matrix~$\Xm$ but is not considered in this paper.

\section{Pareto-Optimal Set for Q-Factor and Radiation Efficiency}
\label{sec:pareto}

It is generally desired to construct an antenna with high efficiency (maximizing~\eqref{eq:eff}) and low Q-factor (minimizing~\eqref{eq:Q}).  By considering the radiation Q-factor~$\Qrad$, we remove the ability to lower Q-factor via a drop in efficiency. Bounds on these quantities for antennas of a certain form-factor can be cast as optimization problems in terms of currents representable in the vector~$\Jm$, where discretization is carried out over the entire design region~\cite{Gustafsson+etal2016a}.

\subsection{Formulation of the Optimization Problems}
\label{sec:pareto:def}

Taking the reciprocal of~\eqref{eq:Q}, the task of minimizing radiation \mbox{Q-factor}~$\Qrad$ can be rewritten as one of maximizing radiated power while maintaining reactive electric and magnetic power below a certain threshold $\Pwc$.  Adding the constraint that losses must also be below a threshold~$\Plc$ yields the optimization problem
\begin{equation}
\begin{aligned}
	& \maximize && \Jm^{\herm}\Rmr\Jm \\
	& \subto &&  \Jm^{\herm}\Xmm\Jm \leq 2\Pwc \\
	& && \Jm^{\herm}\Xme\Jm \leq 2\Pwc \\
	& && \Jm^{\herm}\Rml\Jm \leq 2\Plc. 
\end{aligned}
\label{eq:opt_effQ_tuned}
\end{equation}
Self-resonance can be enforced by only allowing solutions with equal electric and magnetic stored energies.  This amounts to simply changing the inequalities in the energy constraints in the above problem to equalities as
\begin{equation}
\begin{aligned}
	& \maximize && \Jm^{\herm}\Rmr\Jm\\
	& \subto &&  \Jm^{\herm}\Xmm\Jm = 2\Pwc \\
	& && \Jm^{\herm}\Xme\Jm = 2\Pwc \\
	& && \Jm^{\herm}\Rml\Jm \leq 2\Plc.
\end{aligned}
\label{eq:opt_effQ_res}
\end{equation}

\subsection{Relaxation of the Optimization Problems Into Dual Forms}
\label{sec:pareto:dual}

The optimization problems \eqref{eq:opt_effQ_tuned} and \eqref{eq:opt_effQ_res} are not convex and hence not directly solvable.  These problems are, however, indirectly solvable via dual problems \cite{Boyd+Vandenberghe2004}. Here, we present a computationally efficient dual formulation using linear combinations between the constraints in~\eqref{eq:opt_effQ_tuned} and~\eqref{eq:opt_effQ_res}, \ie{},
\begin{equation}\label{eq:convex_effQ_P1}
\begin{aligned}
	& \maximize_{\Jm} && \Jm^{\herm}\Rmr\Jm\\
	& \subto &&  \Jm^{\herm}\Xma\Jm \leq 1,
\end{aligned}	
\end{equation}
where 
\begin{equation}
\Xma = \alpha\nu\Xme+\alpha(1-\nu)\Xmm + (1-\alpha)\Rml
\label{eq:convex_effQ_C1}
\end{equation}
with the right-hand side normalized to unity. In~\eqref{eq:convex_effQ_C1}, the parameter $\alpha\in(0,1)$ is the weight used to sweep the relative priority given to minimizing stored energy or ohmic losses. The boundary value $\alpha=0$ reduces the problem to an optimization of efficiency and $\alpha=1$ represents the minimization of radiation \mbox{Q-factor} presented in~\cite{Capek+etal2017b}. These limiting cases are discussed further in Sec. \ref{sec:comparison}. The Pareto front\footnote{The set of solutions produced by sweeping the parameter $\alpha$ over the interval $[0,1]$.} arising from \eqref{eq:convex_effQ_P1} represents the optimal trade-off between radiation efficiency and radiation Q-factor.

The domain of the dual parameter $\nu$ is $[0,1]$ for~\eqref{eq:opt_effQ_tuned} where the solution current may be tuned by an external reactance.  However this domain is extended to a potentially larger subset of $\R$ in~\eqref{eq:opt_effQ_res} where the solution current is explicitly constrained to be self-resonant. This difference arises from the equalities in~\eqref{eq:opt_effQ_res} that are preserved for arbitrary weights whereas the inequalities in~\eqref{eq:opt_effQ_tuned} are only preserved for non-negative weights. The maximum value $\Jm^{\herm}\Rmr\Jm$ in~\eqref{eq:convex_effQ_P1} is greater or equal to the corresponding values in~\eqref{eq:opt_effQ_tuned} and~\eqref{eq:opt_effQ_res}. Hence,~\eqref{eq:convex_effQ_P1} produces bounds for~\eqref{eq:opt_effQ_tuned} and~\eqref{eq:opt_effQ_res} and its minimization over $\nu$ gives the physical bound.   

In the following section, we discuss the solution of~\eqref{eq:convex_effQ_P1}.  It is important to stress again that, although \eqref{eq:convex_effQ_P1} is an optimization problem, its solution is obtained via deterministic methods based on eigenvalue techniques and simple line search maximization of a function in one variable.

\subsection{Solution of the Optimization Problems}
\label{sec:pareto:solution}

The problem~\eqref{eq:convex_effQ_P1} for a fixed parameter~$\alpha$ is convex~\cite{Boyd+Vandenberghe2004} in $\nu$ and its minimization can be rewritten as maximization of the minimum generalized eigenvalue~\cite{Capek+etal2017b}
\begin{equation}
\maximize_{\nu}\ \min\eig(\Xma,\Rmr),
\label{eq:maxmin}
\end{equation}
where $\eig$ denotes the set of eigenvalues $\{\eigv_{n}\}$ which solves
\begin{equation}
\Xma\Jm_n = \eigv_n\Rmr\Jm_n.
\label{eq:eigv}
\end{equation}
The eigenvalues $\gamma_n$ are all non-negative if $\Xma\succeq\Om$. Thus the maximization of \eqref{eq:maxmin} must occur within the range of $\nu$ within $\R$ to $[\nu_1,\nu_2]$ for which $\Xma\succeq\Om$. The max-min problem~\eqref{eq:maxmin} is easily solved by a line search such as the bisection algorithm~\cite{Boyd+Vandenberghe2004}. Note the formal similarity between~\eqref{eq:eigv} and the eigenvalue problem for characteristic modes~\cite{Harrington+Mautz1972}. The additional maximization over the scalar parameter $\nu$ in~\eqref{eq:maxmin} is solved with a computational cost of a few (order of 10) characteristic mode evaluations using the approach in~\cite{TEAT-7258}.

Solution of \eqref{eq:maxmin} yields the value $\nu_\mathrm{opt}$ which maximizes the minimum eigenvalue $\eigv = \eigv_1$ to  $\eigv_{1,\mathrm{opt}}$.  In solving \eqref{eq:opt_effQ_res} where a self-resonant solution current is required, the solution current~$\Jm_1$ is given directly as the eigenvector corresponding to $\eigv_{1,\mathrm{opt}}$ when $\nu = \nu_\mathrm{opt}$ is used in \eqref{eq:eigv}.  If instead the relaxed problem~\eqref{eq:opt_effQ_tuned} is solved (\ie{}, externally tuned solutions are allowed), the constraint $\nu\in[0,1]$ must be applied.  In this case the solution current $\tilde{\Jm}_1$ and its associated eigenvalue $\tilde{\eigv}_1$ are obtained via \eqref{eq:eigv} using $\nu = \tilde{\nu}_\mathrm{opt}$, where
\begin{equation}
\tilde{\nu}_\mathrm{opt} = \left\{
        \begin{array}{ll}
            \nu_\mathrm{opt} & \quad 0 \leq \nu_\mathrm{opt} \leq 1, \\
            0 & \quad \nu_\mathrm{opt} < 0, \\
            1 & \quad \nu_\mathrm{opt} > 1. \\
        \end{array}
    \right.
    \label{eq:limiting}
\end{equation}

The eigenvalue solving \eqref{eq:maxmin} may be degenerate ~\cite{Capek+etal2017b}. Consider first the case with non-degenerate eigenvalues, a scenario nearly guaranteed for general objects with no represented geometric symmetries \cite{Schab+Bernhard2016}. Under this condition, the derivative of any eigenvalue~\eqref{eq:eigv} with respect to the parameter~$\nu$ is~\cite{Lancaster1964}
\begin{equation}
	\eigv'=\totder{\eigv}{\nu} = \frac{\Jm^{\herm}\Xma'\Jm}{\Jm^{\herm}\Rmr\Jm}
	=-\alpha\frac{\Jm^{\herm}\Xm\Jm}{\Jm^{\herm}\Rmr\Jm},
\label{eq:eigvprim}
\end{equation}
where $\Jm$ is the associated eigenvector.  This shows that the eigenvalue~$\eigv$ is increasing and decreasing if the solution is capacitive ($\Jm^{\herm}\Xm\Jm\leq 0$) and inductive ($\Jm^{\herm}\Xm\Jm\geq 0$), respectively. In particular, it implies that the solution is self-resonant ($\Jm^{\herm}\Xm\Jm= 0$) when $\eigv'=0$. Thus, the solution to \eqref{eq:maxmin} is self-resonant when $\eigv$ is locally maximized.  Note that the restriction on the parameter $\nu$ in the case where external tuning is allowed breaks this condition, as $\nu$ may be limited via \eqref{eq:limiting} to a value of $0$ or $1$ where $\eigv'\neq 0$. This confirms the analysis in~\cite{Capek+etal2017b}, where it is shown that the solution is self-resonant if the optimal value is in the inner region $0<\nu<1$. The self-resonant case with degenerate eigenvalues is identical to~\cite{Capek+etal2017b} where a self-resonant linear combination between the eigenvectors is constructed. However, it is technically more involved, and is left to Appendix~\ref{app:degeneracies}.

\section{Demonstration of the Optimization Procedure}
\label{sec:degen}

\subsection{Pareto-optimality of Q-factor and Radiation Efficiency}
\label{sec:degen:paroptim}

\begin{figure}%
{\centering
\includegraphics[width=1\columnwidth]{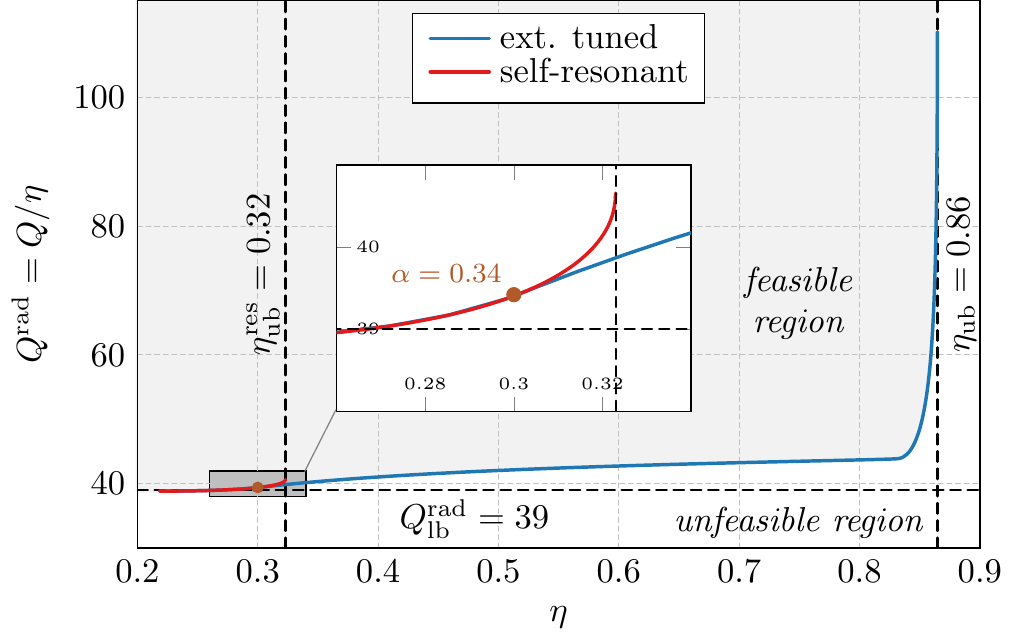}\par}%
\caption{Pareto fronts for externally tuned and self-resonant constraints, \eqref{eq:opt_effQ_tuned} and \eqref{eq:opt_effQ_res}, respectively, calculated via \eqref{eq:convex_effQ_P1} for an L-shaped metallic plate, \mbox{$\Rsurf=1\Ohm$}, \mbox{$ka = 1/2$}, discretized with $1515$~RWG basis functions. An additional marker is added at \mbox{$\alpha=0.34$} for comparison with Fig.~\ref{fig:MOOparameters}. The single-criteria asymptotes~\eqref{eq:eta_ub},~\eqref{eq:eta_ub_res}, and~\eqref{eq:Q_lb} are added as black dashed lines.}
\label{fig:dualproblem}%
\end{figure}

The Pareto front is determined by sweeping the parameter~$\alpha$ in \eqref{eq:maxmin} over the interval~$0<\alpha<1$ and computing the \mbox{Q-factor} and efficiency from the obtained eigenvector. This produces a curve that is interpreted as the optimal trade-off between the minimum \mbox{Q-factor} and maximum efficiency~$\eta$. A Pareto front for a planar L-shaped metallic structure constructed by a rectangle with one quarter removed is depicted in Fig.~\ref{fig:dualproblem}. The structure has side lengths \mbox{$\ell\times\ell/2$}, surface resistance~\mbox{$\Rsurf=1\Ohm$}, electrical size \mbox{$ka = 1/2$}, with~$k$ being the wavenumber and~$a$ being the radius of the smallest sphere circumscribing the radiator, and is discretized with RWG basis functions~\cite{Rao+Wilton+Glisson1982}. Here, the \mbox{Q-factor} is normalized with the efficiency as $\Qrad = Q/\eta$ in order to remove the reduction of \mbox{Q-factor} with decreasing efficiency.

\begin{figure}%
{\centering
\includegraphics[width=1\columnwidth]{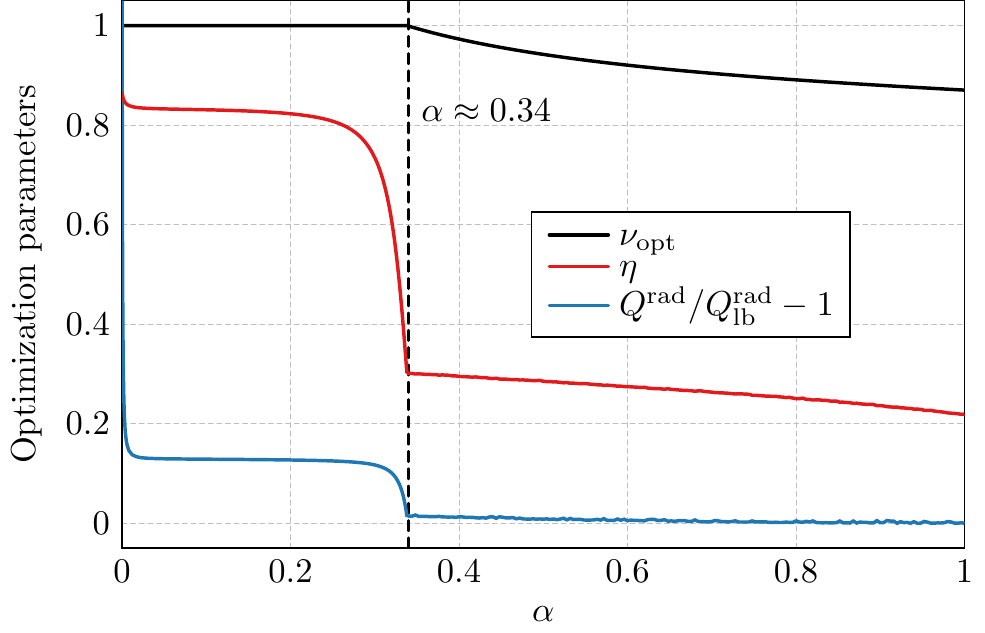}\par}%
\caption{Parameters as functions of the Pareto sweep parameter $\alpha$ from externally tuned solution (blue curve) shown in Fig.~\ref{fig:dualproblem}.  Values of $\alpha$ near $0$ emphasize high efficiency~$\eta$ while values of $\alpha$ near $1$ prioritize low radiation Q-factor~$\Qrad$, \cf{} \eqref{eq:convex_effQ_C1}.}
\label{fig:MOOparameters}%
\end{figure}

The red curve in Fig.~\ref{fig:dualproblem}, representing the Pareto front for a self-resonant current, shows the strict upper bound on self-resonant efficiency, $\eta_\mrm{ub}^\mrm{res}$ (discussed in Section~\ref{sec:comparison}). The Pareto front for the externally tuned case, represented by the blue curve, has a more involved shape. After slightly increasing around the point marked with $\alpha=0.34$, the \mbox{Q-factor} is rather flat with \mbox{$\Qrad\approx 45$} up to \mbox{$\eta\approx 0.85$} where it starts to increase rapidly, reaching \mbox{$\Qrad\approx 115$} at \mbox{$\eta= 0.86$}. It is also observed from Fig.~\ref{fig:MOOparameters} that very small increments in the parameter~$\alpha$ are required to span the Pareto front in this region near \mbox{$\alpha=0$}.  The vertical asymptote at \mbox{$\eta= 0.86$} is approached with \mbox{$\Qrad\rightarrow\infty$} as $\alpha$ is taken closer to zero in finer steps.

From a qualitative point of view, the self-resonant case can be understood as a subset of the externally tuned case. This is confirmed in Fig.~\ref{fig:MOOparameters}, as it can be seen there that the Pareto front valid for externally tuned currents is, in fact, (advantageously) self-resonant for~\mbox{$\alpha \in[0.34,1)$} where \mbox{$0< \nu_\mathrm{opt} < 1$}. The changes in optimized quantities ($Q$, $\eta$) are mostly visible on the left of \mbox{$\alpha = 0.34$}, \ie{}, just before the transition from the self-resonant to the externally tuned region. The price to be paid in the externally tuned region is also seen from Fig.~\ref{fig:MOOparameters} in the blue and red curves for \mbox{$\alpha < 0.34$}. The immediate increase in efficiency implies a question about the practical feasibility of such a high efficiency utilizing external tuning elements because an external compensation of lacking electric/magnetic energy introduces infinitely large losses even for infinitesimally small surface resistivity of that lumped element. Figures~\ref{fig:dualproblem} and \ref{fig:MOOparameters} also suggest that as far as the self-resonant current is concerned, Q-factor and radiation efficiency in this example are almost non-conflicting parameters.

\begin{figure}%
{\centering
\includegraphics[width=1\columnwidth]{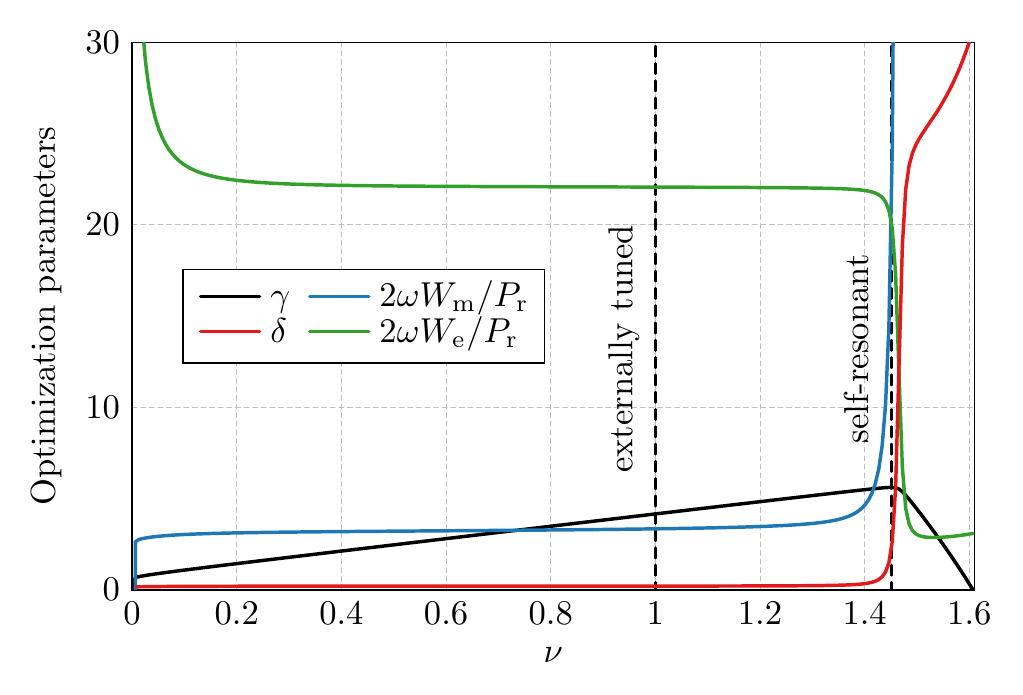}\par}%
\caption{Optimization procedure for a particular choice of \mbox{$\alpha=0.09$}, both for the externally tuned case~\eqref{eq:opt_effQ_tuned} and for the self-resonant case~\eqref{eq:opt_effQ_res}, \mbox{$\Rsurf=1\Ohm$}, \mbox{$ka=1/2$}. The parameter~$\alpha$ has been chosen so that the current is self-resonant ($\gamma$ is maximized) outside~\mbox{$\nu\in\left[0,1\right]$} region. In order to generate the curves, $251$~equidistantly spaced points between \mbox{$\nu_\mathrm{min}=-3.56\cdot 10^{-4}$} and \mbox{$\nu_\mathrm{max}=1.61$}, \ie{}, in region with positive semidefinite operator~$\Xma$ has been used. Notice from Fig.~\ref{fig:MOOparameters} that the self-resonant case is within \mbox{$\nu\in\left[0,1\right]$} only for \mbox{$\alpha>0.34$}.}%
\label{fig:optimprocedure}%
\end{figure}

In order to provide a broader picture of the optimization procedure, the optimized parameters are depicted in Fig.~\ref{fig:optimprocedure} together with the dual parameter~$\nu$ for the particular choice of position on the Pareto front at \mbox{$\alpha=0.09$}, \ie{}, for a position in which the self-resonant solution cannot be attained inside the \mbox{$\nu\in\left[0,1\right]$} interval.

\subsection{Scalability of the Pareto Front with Surface Resistivity~$\Rsurf$}
\label{sec:degen:scalab}

\begin{figure}%
{\centering
\includegraphics[width=1\columnwidth]{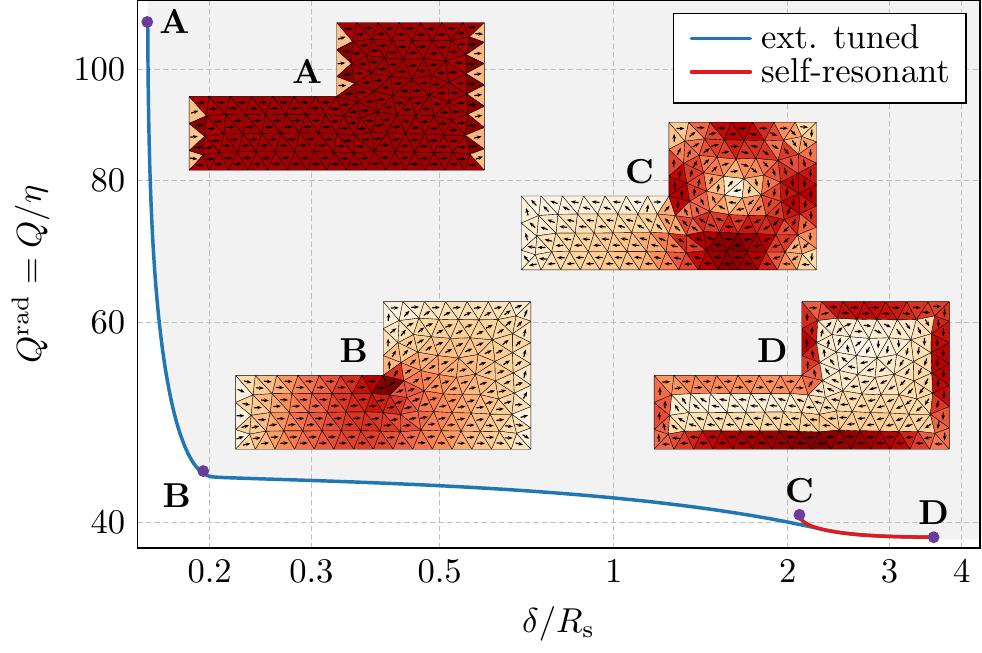}\par}%
\caption{Pareto fronts from Fig.~\ref{fig:dualproblem}, expressed in terms of normalized dissipation factor and radiation Q-factor. Pareto-optimal current distributions are subsampled using $258$~RWG basis functions and shown at four noteworthy values of the parameter~$\alpha$, from left to right: \mbox{$\alpha = \left\{10^{-5}, 0.09, 1\right\}$} for externally tuned solutions denoted by A, B, and D, respectively, and $\alpha = 10^{-5}$ for the self-resonant case denoted by C.}
\label{fig:ParetoDelta}%
\end{figure}

Differentiation of the Rayleigh quotient given by the parameter~$\gamma$ with respect to the Pareto sweep parameter $\alpha$ reads
\begin{equation}
\displaystyle\frac{\mathrm{d}\gamma}{\mathrm{d}\alpha}\Bigg\vert_{\substack{\nu=\nu_\mathrm{opt}\\\alpha=\alpha_0}} =  \Qrad \left(\alpha_0\right) - \delta \left(\alpha_0\right),
\label{eq:pfinpar1}
\end{equation}
which reflects that the trade-off is, in fact, primarily determined by dissipation factor~$\delta$ and radiation quality factor~$\Qrad$, \ie{}, proportionally by the difference between stored energy and ohmic losses, \cf{} \eqref{eq:convex_effQ_C1}. The formula~\eqref{eq:pfinpar1} also clarifies a fact visible in \eqref{eq:convex_effQ_C1}, that the Pareto fronts are scalable with respect to the surface resistance since
\begin{equation}
\frac{\delta_1}{R_{\mrm{s}1}} = \frac{\delta_2}{R_{\mrm{s}2}},
\label{eq:dissipfacttransform}
\end{equation}
thus, the efficiency $\eta_2$ for the Pareto front with resistivity $R_{\mrm{s}2}$ can be obtained from the Pareto front for resistivity $R_{\mrm{s}1}$ and the efficiency~$\eta_1$ with the transformation 
\begin{equation}
\frac{1}{\eta_2} = \frac{R_{\mrm{s}2}}{R_{\mrm{s}1}} \left( \frac{1}{\eta_1} - 1 \right) + 1.
\label{eq:efficiencytransform}
\end{equation}
To demonstrate the effect of the normalization~\eqref{eq:dissipfacttransform}--\eqref{eq:efficiencytransform}, the Pareto front from Fig.~\ref{fig:dualproblem} is redrawn in Fig.~\ref{fig:ParetoDelta} as a function of the dissipation factor~$\delta$, normalized\footnote{Notice that dealing with the normalization~\eqref{eq:dissipfacttransform}--\eqref{eq:efficiencytransform}, one should take care with units as the dissipation factor~$\delta$ is dimensionless and surface resistivity~$\Rsurf$ has units of Ohms ($\Ohm$). Two remedies are possible: either multiply the product $\delta/\Rsurf$ by $1\,\Ohm$ to preserve the dimensionless nature, or to leave $\delta/\Rsurf$ with its units, $\Ohm^{-1} \equiv \mathrm{S}$, keeping in the mind that~$\Rsurf$ is used as a scaling parameter only. The later option is used throughout this paper.} by used surface resistivity~$\Rsurf$, and radiation Q-factor~$\Qrad$. The insets depict Pareto optimal surface current densities for four distinct positions on the Pareto fronts. The minimum dissipation factor for the externally tuned case~A is realized by a constant current density as predicted by~\cite{Shahpari+Thiel2016}. The solution~D for minimum radiation Q-factor is common for both self- and externally tuned currents and consists of a mixture of dominant capacitive and inductive modes~\cite{Capek+Jelinek2016,Capek+etal2017b}. The self-resonant solution~C with minimum dissipation factor is realized by a loop current with smoothly varying amplitude, which is in contrast with case~D, where the current is predominantly distributed along the boundary edges. The last depicted inset~B shows the current density in the transition region between low and high Q-factor and it resembles a sine current on a dipole-type antenna. To remove the dependence on electrical size, the x- and y- axes can further be normalized by~$\left(ka\right)^4$ and~$\left(ka\right)^3$, respectively, which makes it possible to compare electrically small antennas irrespective of their electrical size and surface resistivity.

A closer look at~\eqref{eq:pfinpar1} reveals two notable properties. First, integrating~\eqref{eq:pfinpar1} and substituting the limiting values of Pareto sweeping parameter $\alpha$ yields
\begin{equation}
\label{eq:pfinpar3}
\int_{0}^{1} \frac{\mathrm{d}\gamma}{\mathrm{d}\alpha}\Bigg\vert_{\nu_\mathrm{opt}} \, \mathrm{d}\alpha = \Qradlb - \delta_\mrm{lb}
\end{equation}
\ie{}, the difference between the solutions of individual single-criteria problems. Second, the differentiation of~\eqref{eq:pfinpar1} readily confirms that all non-dominated solutions (members of the Pareto front) have negative second derivatives of~$\gamma \left(\alpha\right)$. Notice, however, that the contrary (all solutions with second derivative negative are members of the Pareto front) is not supported.

\section{Comparison With Limiting Cases}\label{sec:comparison}

The Pareto fronts are compared with their limiting cases where some of the constraints in~\eqref{eq:opt_effQ_tuned} and~\eqref{eq:opt_effQ_res} are removed.

\subsection{General Cases}
\label{sec:comparison:general}

The upper bound on the radiation efficiency~$\eta$ is determined from the minimum dissipation factor that results in the eigenvalue problem~\cite{Jelinek+Capek2017,Harrington1960}
\begin{equation}
\left(\eta_\mrm{ub}\right)^{-1} - 1 = \delta_\mrm{lb} = \min \eig(\Rml,\Rmr).
\label{eq:eta_ub}
\end{equation} 
In comparison, the upper bound on efficiency $\eta_\mathrm{ub}^\mathrm{res}$ attainable by a self-resonant current follows from \eqref{eq:opt_effQ_res} by adding the redundant constraint of self-resonance, \mbox{$\Jm^{\herm}\Xm\Jm=0$}, and then removing the constraints on the actual values of electric and magnetic powers, \ie{},~\cite{Jelinek+Capek2017}
\begin{equation}
\begin{aligned}
	& \maximize && \Jm^{\herm}\Rmr\Jm\\
	& \subto &&  \Jm^{\herm}\Rml\Jm = 1 \\
	& && \Jm^{\herm}\Xm\Jm = 0. 
\end{aligned}
\label{eq:opt_eff_res}
\end{equation}
This reformulation represents the boundary value of \mbox{$\alpha = 0$} for the Pareto analysis of \eqref{eq:opt_effQ_res} and can be relaxed to the dual problem 
\begin{equation}\label{eq:opt_eff_res_dual}
\begin{aligned}
	& \minimize_{\nu}\maximize_{\Jm} && \Jm^{\herm}\Rmr\Jm\\
	& \subto &&  \Jm^{\herm}(\nu\Xm+\Rml)\Jm = 1 
\end{aligned}	
\end{equation}
analogously to the analysis of~\eqref{eq:convex_effQ_P1}. The solution is
\begin{equation}
\left(\eta_\mrm{ub}^{\mrm{res}}\right)^{-1} - 1 = \delta_\mrm{lb}^\mrm{res} = \max_{\nu\in\R}\min \eig(\nu\Xm+\Rml,\Rmr)
\label{eq:eta_ub_res}
\end{equation}
while~\eqref{eq:eigvprim} shows that the constraints in~\eqref{eq:opt_eff_res} are satisfied at the optimal value for non-degenerate eigenvalues. A self-resonant solution is constructed~\cite{Capek+etal2017b} for the corresponding degenerate case showing that~\eqref{eq:eta_ub_res} solves~\eqref{eq:opt_eff_res}. This formulation~\eqref{eq:eta_ub_res} for the lower bound on the dissipation factor~$\delta$ is concave and hence easily solvable. With \mbox{$\Rml = \Rsurf\boldsymbol{\Psi}$}, we have the solution for all values of surface resistance according to~\eqref{eq:efficiencytransform} and it is hence sufficient to solve~\eqref{eq:eta_ub_res} for $\Rsurf=1\Ohm$. The final limiting case, the lower bound on the radiation Q-factor, is determined as~\cite{Capek+etal2017b}
\begin{equation}
\Qradlb  =\max_{0\leq\nu\leq1}\min \eig\left(\nu\Xme+(1-\nu)\Xmm,\Rmr\right).
\label{eq:Q_lb}
\end{equation}

Here we continue studying the example given in Section~\ref{sec:pareto} with the addition of these three limiting cases. The Pareto-fronts for the tuned~\eqref{eq:opt_effQ_tuned} and self-resonant~\eqref{eq:opt_effQ_res} cases overlap for low efficiencies where their \mbox{Q-factors} are close to the lower bound~$\Qradlb$, see the magnified part in Fig.~\ref{fig:dualproblem}. The \mbox{Q-factor} for the self-resonant case increases slightly and reaches the upper bound in the efficiency $\eta_\mrm{ub}^{\mrm{res}}$~given by \eqref{eq:eta_ub_res}. We now identify the vertical asymptote in the externally tuned Pareto front as the absolute maximum value of efficiency \mbox{$\eta_\mrm{ub}$} given by \eqref{eq:eta_ub}. The upper bound $\eta_{\mrm{a}}$ proposed in~\cite{Shahpari+Thiel2016} solely based in the surface area, resistivity, and frequency gives the same value as the bound \mbox{$\eta_\mrm{ub}$}. These bounds with their approximately constant current distribution (see \mbox{$\alpha = 10^{-5}$} point in Fig.~\ref{fig:dualproblem}) compared with the closer to cosine-shaped optimal current distribution~\cite{Gustafsson+etal2012a} for minimum \mbox{Q-factor} of electric dipole radiators explain the rapid increase of the~$\Qrad$. Note that a cosine-shaped current distribution doubles the ohmic losses compared with the idealized constant current for a rectangle and that the corresponding dissipation factor increases as \mbox{$\pi^2/8\approx 1.23$} for small structures~\cite{Shahpari+Thiel2016}. The \mbox{Q-factor} increases rapidly as the charge accumulates at the edges when the approximately constant current distribution is forced to vanish at the edge.     

\subsection{Pareto Optimality for Spherical Shell}
\label{sec:comparison:sphshell}

A geometry with interesting behavior in terms of these bounds is the spherical shell, for which the optimal currents and bounds for Q-factor and efficiency are known analytically~\cite{Chu1948,Thal2006,Losenicky+etal2018}. As a single current is optimal in both self-resonant radiation Q-factor and efficiency, the self-resonant Pareto front is represented by one point with an optimal current consisting of a mixture of dominant TM and TE modes
\begin{equation}
\Jm = \Jm_\mathrm{TM10} + \xi \left( ka \right) \Jm_\mathrm{TE10},
\label{eq:sphShell1}
\end{equation}
with the coupling constant~\cite{Capek+Jelinek2016, Losenicky+etal2018}
\begin{equation}
\xi_\mrm{opt} = \sqrt{- \displaystyle\frac{1 - ka \displaystyle\frac{\mathrm{y}_0 \left(ka\right)}{\mathrm{y}_1 \left(ka\right)}}{1 - ka \displaystyle\frac{\mathrm{j}_0 \left(ka\right)}{\mathrm{j}_1 \left(ka\right)}}} \mathrm{e}^{\mathrm{j} \varphi},
\label{eq:sphShell2}
\end{equation}
and $\mathrm{j}_n \left(ka\right)$, $\mathrm{y}_n \left(ka\right)$ spherical Bessel functions of the $n$th order and of the first and second kind~\cite{Arfken+Weber2005}, respectively. Currents lying on the externally tuned Pareto front follow the form of~\eqref{eq:sphShell1} and sweep continuously between~$\xi = 0$ and \mbox{$\xi = \xi_\mrm{opt}$}. The boundary values delimit the interval between the externally tuned maximum radiation efficiency ($\xi=0$) and naturally self-resonant minimum Q-factor ($\xi = \xi_\mrm{opt}$) current distributions realizable on a spherical shell.

Knowledge of \eqref{eq:sphShell1}, the interval of $\xi$, and orthogonality of spherical harmonics with respect to the operators~$\Rmr$, $\Xmm$, and $\Xme$ makes it possible to determine the Pareto front analytically. Suppose the set of spherical harmonics is normalized to unitary radiated power, \mbox{$P_\mrm{r, TM10} = P_\mrm{r, TE10} = 1$}, \ie{} characteristic modes of the spherical shell. Then the radiation Q-factor is equal to
\begin{equation}
\Qrad = \frac{\omega \max \left\{W_\mrm{m,TM10+TE10}, W_\mrm{e,TM10+TE10} \right\}}{1 + \xi^2},
\label{eq:sphShell3}
\end{equation}
where
\begin{align}
W_\mrm{m,TM10+TE10} &= W_\mrm{m,TM10} + \xi^2 W_\mrm{m,TE10}, \\
W_\mrm{e,TM10+TE10} &= W_\mrm{e,TM10} + \xi^2 W_\mrm{e,TE10},
\label{eq:sphShell4}
\end{align}
and where the individual modal energies are evaluated in~\cite{Hansen+Collin2009}. The resulting dissipation factors are equal to~\cite{Losenicky+etal2018}
\begin{equation}
\delta = \frac{P_\mrm{\Omega, TM10} + \xi^2 P_\mrm{\Omega, TE10}}{1 + \xi^2}.
\label{eq:sphShell5}
\end{equation}

\begin{figure}%
{\centering
\includegraphics[width=1\columnwidth]{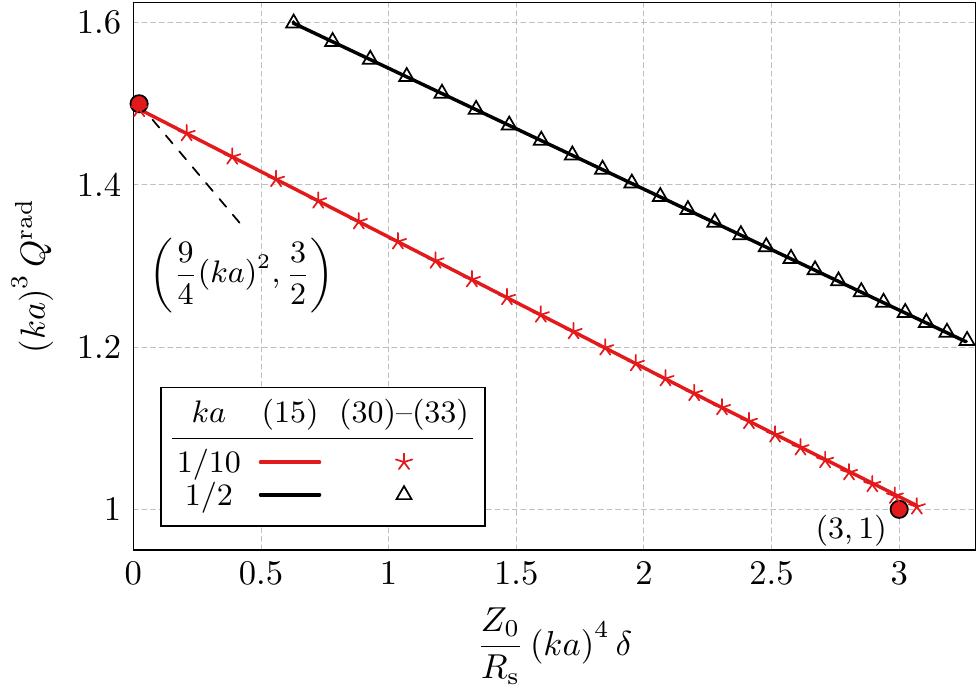}\par}%
\caption{Pareto fronts for externally tuned currents on a spherical shell of radius~$a$. The current is self-resonant for both Pareto fronts at the extreme points on the right side, where the coupling coefficient~\mbox{$\xi = \xi_\mrm{opt}$}. The numerical data were obtained from a spherical shell discretized with $1215$~RWG basis functions. The parameter~$\xi^2$ is swept linearly in interval~\mbox{$\left[0, \xi_\mrm{opt}^2\right]$} with $25$ points. The additional circular markers denote the limiting cases \eqref{eq:sphShell6}, \eqref{eq:sphShell7}.}
\label{fig:SphShellPareto}%
\end{figure}

The normalized Pareto fronts for a spherical shell of electrical sizes \mbox{$ka = \left\{1/10, 1/2\right\}$} are depicted in Fig.~\ref{fig:SphShellPareto}. Solid lines are solutions to~\eqref{eq:sphShell3}, \eqref{eq:sphShell5}, dashed lines represent solutions to~\eqref{eq:maxmin}. Since the denominators of~\eqref{eq:sphShell3} and~\eqref{eq:sphShell5} diminish in their mutual trade-off, it is obvious that the Pareto front is a straight line. 

For small $ka$ ($ka\ll 1$), the asymptotic values for the end points are known as~\cite{Thal2006, Losenicky+etal2018}
\begin{align}
\label{eq:sphShell6}
(ka)^3 \Qradlb = 1 \quad &\textrm{and} \quad (ka)^3 \Qrad_\mrm{TM10} = \frac{3}{2}, \\
\label{eq:sphShell7}
\frac{Z_0}{\Rsurf}(ka)^4 \delta = 3 \quad &\textrm{and} \quad \frac{Z_0}{\Rsurf} (ka)^4 \delta_\mrm{lb} = \frac{9}{4} (ka)^2,
\end{align}
respectively. Two extrema points given by~\eqref{eq:sphShell6} and~\eqref{eq:sphShell7} are depicted in Fig.~\ref{fig:SphShellPareto} by circular markers.

\subsection{Pareto Optimality for Layered Prolate Spheroid}
\label{sec:comparison:spheriod}

\begin{figure}%
{\centering
\includegraphics[width=1\columnwidth]{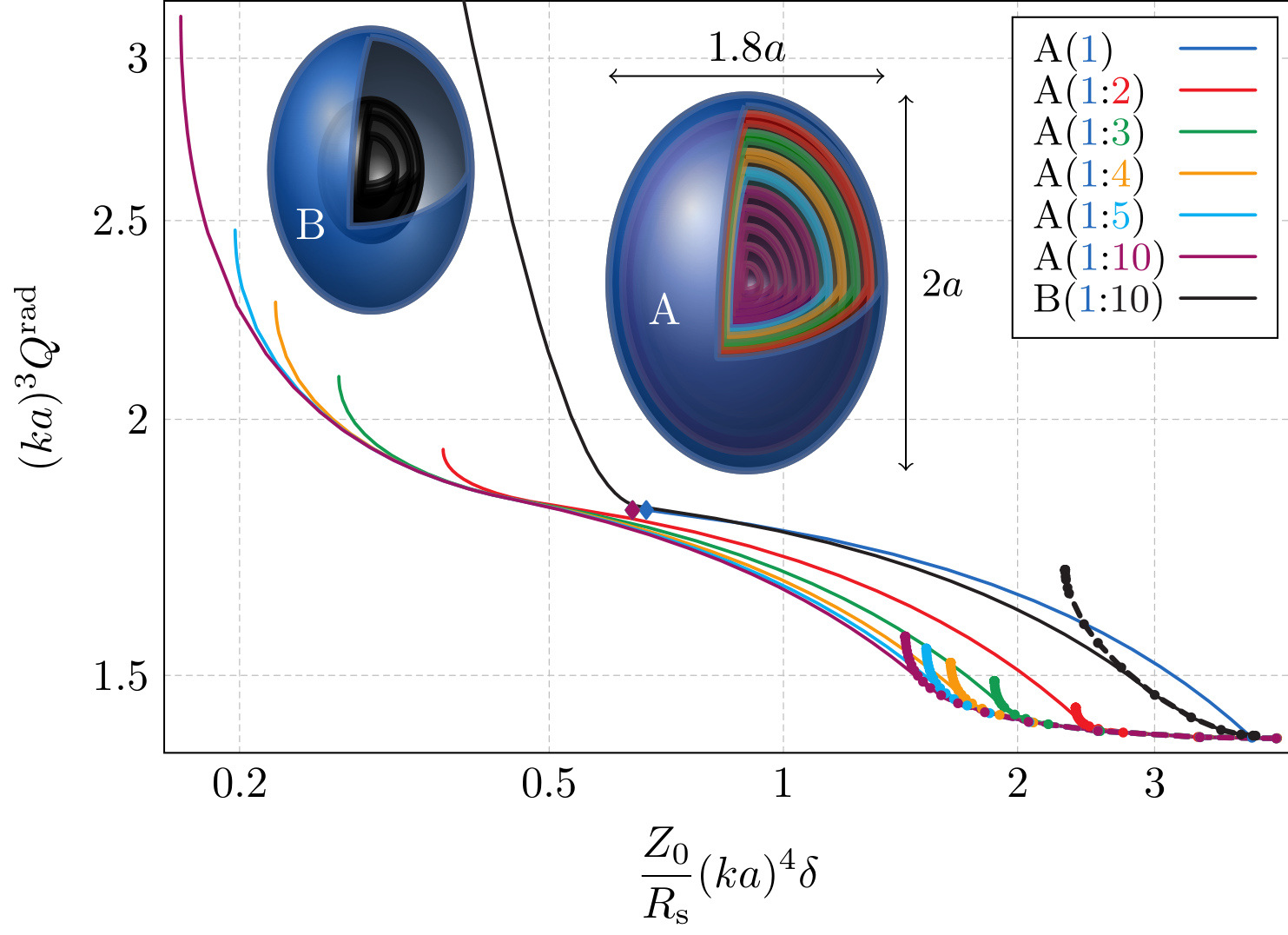}\par}%
\caption{Pareto curves for layered prolate spheroids with semi axes $a$ and $0.9a$ for $ka=0.5$. Solid and dotted curves for tuned~\eqref{eq:opt_effQ_tuned} and self resonant~\eqref{eq:opt_effQ_res}, respectively. The layers have resistivity~$\Rsurf$ and semi axes scaled $(11-m)/10$, where $m\in\{1,...,n\}$ for the curve labeled $1-n$. The black curves  correspond to the B(1:10) case with the inner~$9$ layers having semi axis equidistantly scaled between~$0.25$ and~$0.5$.} 
\label{fig:FilledSpheroid}%
\end{figure}

Pareto curves for a layered prolate spheroid with semi axes $a$ and $0.9a$ and electrical size $ka=1/2$ are depicted in Fig.~\ref{fig:FilledSpheroid}. The layers have surface resistivity $\Rsurf$ and form spheroids with axes scaled by $(11-m)/10$ with $m\in\{1,...,n\}$ for curves labeled $\{1,...,n\}$. The figure also contains the Pareto front for the case with of nine inner layers placed on spheroids with axes scaled between $0.25$ and $0.5$ of the radial distance. 

The Pareto front for the self-resonant~\eqref{eq:opt_effQ_res} single layer case is close to a single point, approaching the solution obtained from a sphere~\eqref{eq:sphShell1}. Its normalized dissipation and Q-factors are around $4$ and $1.4$, respectively, which are slightly higher than for the sphere~\eqref{eq:sphShell6}. The Pareto front for tuned case~\eqref{eq:opt_effQ_tuned} traverses from the self-resonant mixed TE/TM mode~\eqref{eq:sphShell1} to the TM mode around $(0.7,1.8)$ which is also obtained by evaluating the minimum Q-factor for TM radiation~\cite{Capek+etal2017b} and marked with diamonds in Fig.~\ref{fig:FilledSpheroid}.

Filling the spheroid highlights some differences between the lower bounds on the dissipation and Q-factor. Here, internal layers are added, rather than a homogeneous filling~\cite{Karlsson2013,Fujita+Shirai2015}, to simplify the comparison with the surface currents in this paper. We also note that there is a trivial scaling of the sheet resistivity~\eqref{eq:sheet} as two closely spaced layers can be replaced by a single layer with the double thickness and hence half the sheet resistivity. It is seen that the Pareto front extends as the number of inner layers are included. The effect is largest for the layers close to surface and the contributions from the layers close to the center are negligible as seen from the small difference between the A(1:5) and A(1:10) cases. Through the introduction of additional internal layers, the dissipation factor reduces while the lower bound on radiation Q-factor is unaffected by the possible sources in the inner region~\cite{Gustafsson+etal2012a}. This shows that it is possible to utilize the interior region of an object to decrease losses and that volumes should be considered in physical bounds on dissipation factor.     

One possibility for using the inner region is to tune the dominant electric energy from the outer shell magnetic energy in the inner region. This is illustrated by the case denoted B(1:10) that consists of nine layers placed between 0.25 and 0.5 of the radial distance. The lower bound for the self-resonant case traverses from the mixed mode solution to a dominant TM mode.

\section{Limit of electrically small antennas}\label{sec:sizelimits}

The low-frequency (electrically small) limit offers explicit solutions for many antenna problems~\cite{Yaghjian+Stuart2010,Gustafsson+etal2012a,Yaghjian+etal2013,Jonsson+Gustafsson2015} and can be used to compute the dominant components of the small size expansions~\eqref{eq:sphShell6} and~\eqref{eq:sphShell7} for arbitrary shaped objects. Under the low-frequency expansion of a current density \mbox{$\Jv\sim\Jv_0+k\Jv_1$} (as $k\to 0$ where $\nabla\cdot\Jv_0=0$ and $\nabla\times\Jv_1=\Ov$), the optimization problem for minimum Q-factor is separated into its electric and magnetic parts~\cite{Gustafsson+etal2012a}. The stored electric and magnetic energies themselves have low-frequency asymptotic expansions\footnote{The components $\Jm_0$ and $\Jm_1$ may be represented by lower dimension bases, however for notational simplicity we assume here that they are expressed in the same basis as a general current vector $\Jm$.}
\begin{equation}
  \We=\frac{1}{4\omega}\Jm_1^{\herm}\Xme\Jm_1
  \sim\frac{1}{4}\Jm_1^{\herm}\Cmi\Jm_1
\label{eq:lowFreqWe}
\end{equation}
and
\begin{equation}
  \Wm=\frac{1}{4\omega}\Jm_0^{\herm}\Xmm\Jm_0
  \sim\frac{1}{4}\Jm_0^{\herm}\Lm\Jm_0,
\label{eq:lowFreqWm}
\end{equation}
respectively, where $\Lm$ and $\Cmi$ are defined by the impedance matrix $\Zm=\ju k\Lm-\ju\Cmi/k$. The radiated power and dissipated power in ohmic losses can be written
\begin{equation}
  \Prad \sim \frac{1}{2}k^4(\Jm_0^{\herm}\Rmrm\Jm_0+\Jm_1^{\herm}\Rmre\Jm_1)
  \text{ and }
  \Pl \sim \frac{1}{2}\Jm_0^{\herm}\Rml\Jm_0,
\label{eq:Rem}
\end{equation}
respectively, with the explicit representations of the matrices, \eg{}, given~\cite[Eq. (3.9)]{Jonsson+Gustafsson2015}. Inserting the expansions in~\eqref{eq:Q_lb} shows that the eigenvalue problem  separates into TM and TE eigenvalue problems
\begin{equation}
  k^3 \QradlbTM
	=\min\eig_{\nabla\times}(\Cmi,\Rmre) 
  = \frac{6\pi}{\max\eig\gammame} 
  \label{eq:eigCRr}
\end{equation}
and 
\begin{equation}  
  k^3 \QradlbTE
	=\min\eig_{\nabla\cdot}(\Lm,\Rmrm)
  = \frac{6\pi}{\max\eig\gammamm} 
  \label{eq:eigLRr}
\end{equation}
which have explicit solutions expressed in the electric and magnetic polarizability dyadics~$\gammame$ and~$\gammamm$, respectively, see also~\cite{Gustafsson+etal2007a, Gustafsson+etal2012a,Yaghjian+etal2013,Jonsson+Gustafsson2015,Gustafsson+etal2015b} for alternative derivations. In \eqref{eq:eigCRr} and \eqref{eq:eigLRr}, $\eig_{\nabla\times}$ and $\eig_{\nabla\cdot}$ denote solving the eigenvalue problem while admitting only curl- and divergence-free solutions, respectively. Note that this notation assumes some regularity on the current density and generalization are needed for surface currents and complex topologies. 
The lower bound on the Q-factor~\eqref{eq:Q_lb} for combined TE and TM modes has the explicit representation
\begin{equation}
\Qradlb=\frac{6\pi}{k^3(\max\eig\gammame+\max\eig\gammamm)}.
\label{eq:Qlbgamma}
\end{equation}

The low-frequency expansion transforms the optimization for the minimal self-resonant dissipation factor~\eqref{eq:opt_eff_res} to
\begin{equation}
\begin{aligned}
	& \maximize && \Jm_0^{\herm}\Rmrm\Jm_0+\Jm_1^{\herm}\Rmre\Jm_1\\
	& \subto &&  \Jm_0^{\herm}\Rml\Jm_0 = 1 \\
	& && \Jm_0^{\herm}\Lm\Jm_0 = \Jm_1^{\herm}\Cmi\Jm_1,
\end{aligned}
\label{eq:opt_eff_res_lowf1}
\end{equation}
where the electric and magnetic part separates. The electric part is maximized in~\eqref{eq:eigCRr} simplifying the optimization problem to
\begin{equation}
\begin{aligned}
	& \maximize && \Jm_0^{\herm}\Rmrm\Jm_0+\beta\\
	& \subto &&  \Jm_0^{\herm}\Rml\Jm_0 = 1 \\
	& && \Jm_0^{\herm}\Lm\Jm_0 = 6\pi\beta/\gamma,
\end{aligned}
\label{eq:opt_eff_res_lowf2}
\end{equation}
where $\gamma=\max\eig\gammame$.  The above problem is thus reduced to a maximization over $\Jm_0$ and the parameter $\beta$. This optimization is solved analogously to~\eqref{eq:opt_eff_res}. The $k^4$ scaling of the radiated power in~\eqref{eq:Rem} shows that the dissipation factor scales as $k^{-4}$.  . The electric part (charge) is hence identical for minimum Q-factors of the electric dipole (TM)~\eqref{eq:eigCRr}, and combined mode (TM+TE)~\eqref{eq:Qlbgamma}, and self resonant maximal efficiency cases~\eqref{eq:opt_eff_res_lowf2}. Maximization of~\eqref{eq:opt_eff_res_lowf2} includes both dissipation in ohmic losses $\Rml$ and magnetic radiation $\Rmrm$. Assuming an electric dipole radiator, \ie{},   
negligible magnetic radiation in~\eqref{eq:opt_eff_res_lowf2}, gives an optimization problem for $\Jm_0$ in~\eqref{eq:opt_eff_res_lowf2} as the maximum Q-factor of an inductor 
\begin{equation}
  \frac{\QL}{k}
  =\max\frac{\Jm_0^{\herm}\Lm\Jm_0}{\Jm_0^{\herm}\Rml\Jm_0}
	=\max\eig_{\nabla\cdot}(\Lm,\Rml).
\label{eq:QL}
\end{equation}
Inserting~\eqref{eq:QL} into~\eqref{eq:opt_eff_res_lowf2} yields the minimal TM dissipation factor as the quotient between the TM radiation and inductor Q-factors 
 \begin{equation}
  \delta_{\mrm{lb,TM}} 
  =\frac{\min\eig_{\nabla\times}(\Cmi,\Rmre)}{k^4 \max\eig_{\nabla\cdot}(\Lm,\Rml)}
  = \frac{6\pi}{k^3\gamma \QL}
  = \frac{\QradlbTM}{\QL}.
\label{eq:dfTM}
\end{equation}
This solution can be interpreted as stored energy from a lossless electric dipole radiators tuned with lossy inductive currents. 

\begin{figure}%
{\centering
\includegraphics[width=1\columnwidth]{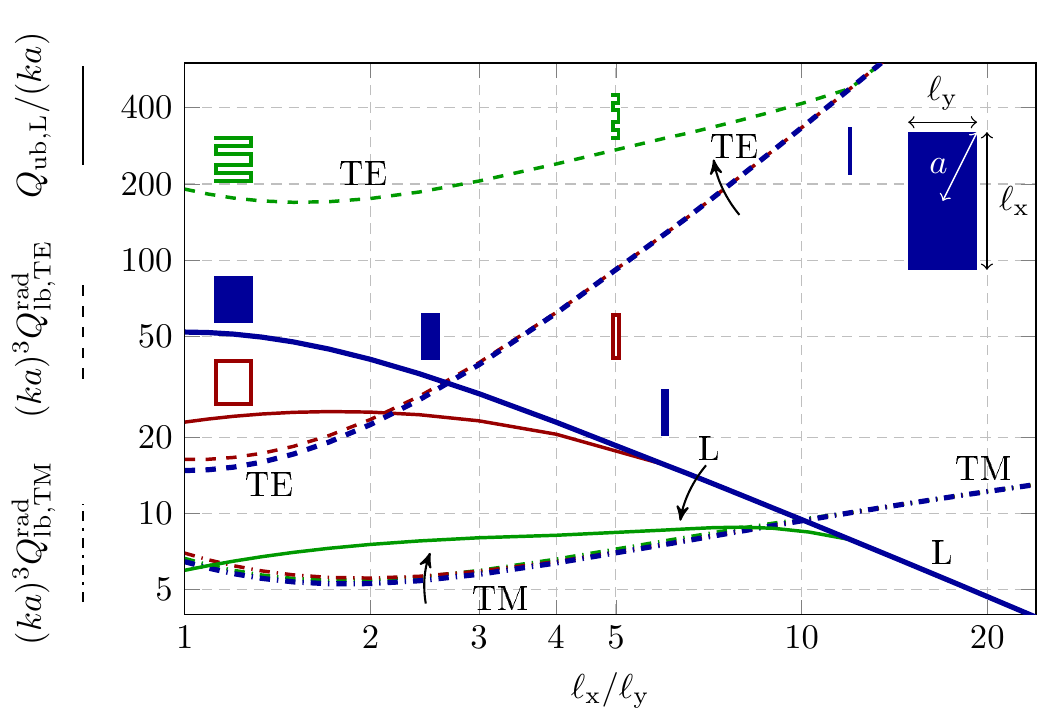}\par}%
\caption{Normalized Q-factors for electric dipole antennas $\QradlbTM(ka)^3$~\eqref{eq:eigCRr} (dot dashed), magnetic dipole $\QradlbTE(ka)^3$~\eqref{eq:eigLRr} (dashed), and inductors $\QL/(ka)$~\eqref{eq:QL} (solid) confined to a rectangular region with side lengths $\ellx$ and $\elly$. Effects on the Q-factors from removal of the inner region (loop) and slots (meandering) are also included.}
\label{fig:QtmQLrec}%
\end{figure}

Q-factor bounds $\QradlbTM$, $\QradlbTE$, and $\QL$ are depicted in Fig.~\ref{fig:QtmQLrec} for planar rectangles with aspect rations $\ellx/\elly$ and two antenna geometries circumscribed by the rectangles. The loops and meander lines have strip widths $\ellx/12$ which imply that the loop and meander line reduce to the rectangle for $\ellx\geq 6\elly$ and $\ellx\geq 12\elly$, respectively.   
The TM and TE Q-factors in dashed dotted and dashed curves, respectively, are normalized with $(ka)^3$ and related to the polarizability dyadic, see~\eqref{eq:eigCRr} and~\eqref{eq:eigLRr}. 

The loop and meander shapes are formed be removing parts of the rectangle and are hence suboptimal in~\eqref{eq:eigCRr}, \eqref{eq:eigLRr}, and~\eqref{eq:QL} compared to the solution for the circumscribing rectangle. Suboptimal solutions to the minimization in ~\eqref{eq:eigCRr} and~\eqref{eq:eigLRr} imply increased Q-factors $\QradTM$ and $\QradTE$.  Similarly, suboptimal solutions imply reduction to the maximization of $\QL$ in~\eqref{eq:QL}, see arrows in Fig.~\ref{fig:QtmQLrec}. This is a similar conclusion as drawn from the variational bounds for the polarizability dyadics~\cite{Gustafsson+etal2007a}. However, the electric polarizability is only marginally decreased as the removed parts are located in the inner region of the rectangle~\cite{Gustafsson+etal2015b} and hence are the corresponding Q-factors approximately the same for the antenna geometries as for the circumscribing rectangle. 
Contrary, the TE $\QradlbTE$ and inductive $\QL$ Q-factor change rapidly as inner parts are removed. This can be attributed to elimination of loop currents when such regions are removed or cut by slots. 

The small increases of $\QradlbTE$ for the loop compared with the rectangle indicate that the magnetic polarizability is dominated by the loop area. The corresponding much larger decrease in $\QL$ shows that the inductance can benefit more from the inner parts, \cf{} with the C and D cases in Fig.~\ref{fig:ParetoDelta}. The approximate factor of two between $\QradlbTM$ and $\QradlbTE$ for $\ellx=\elly$ indicate a potential $66\%$ reduction for the combined TM-TE mode~\eqref{eq:Q_lb}. This reduction however vanished rapidly with increasing aspect ratio as the quotient $\QradlbTM/\QradlbTE\gg1$ for $\ellx\geq 2\elly$. 

The impact from the slots in the meander lines is much larger than for the loops as the slots hinder large loop currents spanning the entire surface. Here, $\QradlbTE$ increases and $\QL$ decreases close to a factor of 10. These large changes in $\QradlbTE$ and $\QL$ with respect to removal of geometry compared to the small changes of $\QradlbTM$ partly explains the increased difficulty to design optimal antennas for $\Qrad_{\mrm{TE}}$, $\Qrad$, and efficiency compared to $\Qrad_{\mrm{TM}}$.  
The solutions to~\eqref{eq:eigCRr} and~\eqref{eq:QL} are qualitatively different for surface and volumetric regions $\reg$. Minimization of the Q-factors~\eqref{eq:eigCRr} and~\eqref{eq:eigLRr} have surface charge and surface current distributions, respectively~\cite{Gustafsson+etal2012a}. The maximization of the inductor Q-factor~\eqref{eq:QL} has instead a volumetric current distribution, \cf\ the filled spheroid in Fig.~\ref{fig:FilledSpheroid}.

\section{Controllable Region Constraints}\label{sec:subregion}

Restricting control of currents to a subregion of a structure is equivalent to adding a linear constraint \mbox{$\Jm_{\mrm{G}}=\Tm\Jm_{\mrm{A}}$} to~\eqref{eq:opt_effQ_tuned} and~\eqref{eq:opt_effQ_res}, see~\cite{Gustafsson+Nordebo2013,Capek+etal2017b} for details. This restriction shifts the Pareto front into the region of feasible solutions; that is, less control leads to stricter bounds in both Q-factor and efficiency. 

\begin{figure}%
{\centering
\includegraphics[width=1\columnwidth]{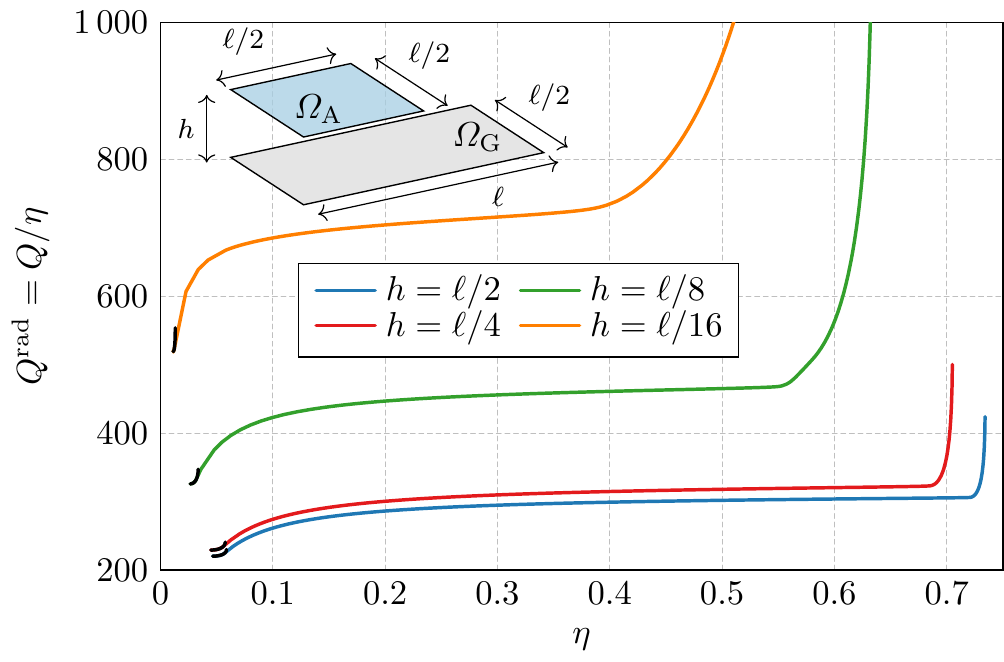}\par}%
\caption{The Pareto fronts for square patch with controllable sources~$\varOmega_\mathrm{A}$ placed in various heights~$h$ over a rectangular ground plane~$\varOmega_\mathrm{G}$, see the inset. The Pareto-optimal currents are considered to be tuned to resonance by external elements, \ie{}, problem~\eqref{eq:opt_effQ_tuned} was solved at \mbox{$ka=1/2$} for \mbox{$\Rsurf=1\Ohm$}. The self-resonance Pareto fronts, fulfilling~\eqref{eq:opt_effQ_res} and depicted by solid black lines, are added only for the completeness as they are of extremely limited scope. In all cases, the same mesh grid was used with $202$~RWG elements in the controllable region and~$422$~RWG elements on the ground plane.}%
\label{fig:subdomain}%
\end{figure}

Here we examine a simple example of a calculation involving this class of linear constraints imposed by incomplete control of currents within a design region.  The sample problem, shown in Fig.~\ref{fig:subdomain}, consists of an antenna design region~$\varOmega_\mathrm{A}$ where currents represented by~$\Jm_\mrm{A}$ are fully controllable.  Below this region is a larger ground plane~$\varOmega_\mathrm{G}$ where currents represented by~$\Jm_\mrm{G}$ are induced according to the linear relation~\mbox{$\Jm_{\mrm{G}}=\Tm\Jm_{\mrm{A}}$}.  

The height~$h$ of the controllable region over the ground plane is adjusted and the resulting data in Fig.~\ref{fig:subdomain} demonstrate how this distance affects the Pareto fronts.  Both extremal points on the Pareto front degrade with decreasing height within the range studied here, \ie{}, minimum radiation Q-factor increases while maximum efficiency decreases.  Additionally, the interior of the Pareto fronts become smoother with decreasing height, indicating that the trade-off between these quantities becomes more pronounced as the controllable region is brought closer to the ground plane.  Specifically, for the largest value of~$h$ studied here, an efficiency marginally below the maximum ($71\%$ vs. $73\%$) can be achieved by a current with radiation Q-factor only $40\%$ above the minimum.  However, at the smallest value of~$h$, the same relative sacrifice in radiation Q-factor leads to an efficiency well below the optimal value ($36\%$ vs. $59\%$).  These results can further be compared to the high cost of approaching the absolute minimum possible Q-factor, which in all cases studied here dictates a low efficiency on the order of $5\%$ or less.

\begin{figure}%
{\centering
\includegraphics[width=1\columnwidth]{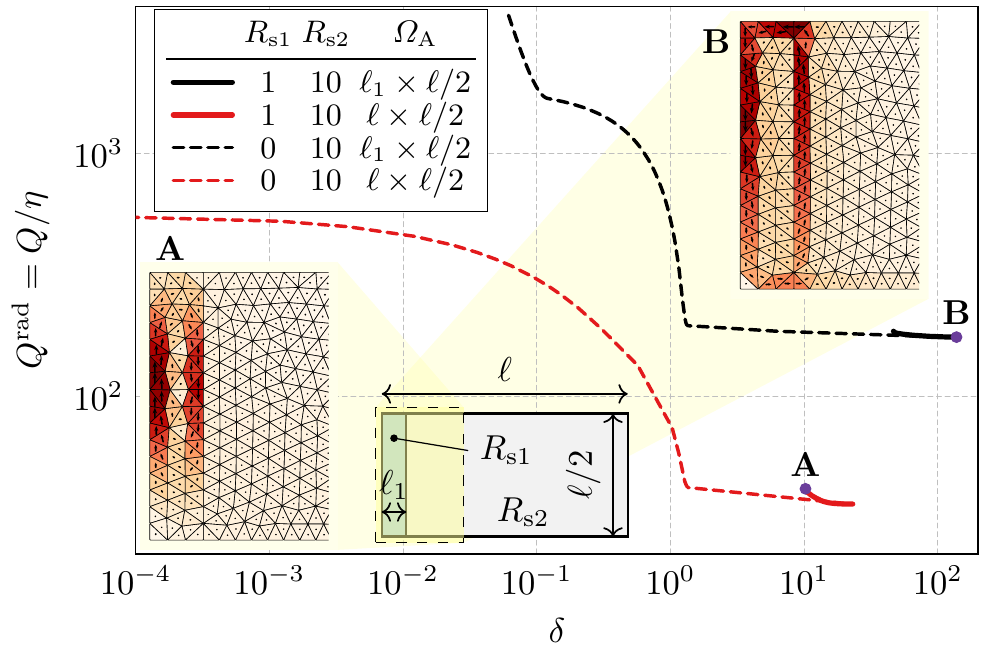}\par}%
\caption{Pareto fronts calculated with the self-resonant constraint. The rectangular plate of \mbox{$\ell\times\ell/2$} dimensions, \mbox{$ka=1/2$}, with 1478~RWG basis functions was divided into two regions of different surface resistivity, $R_\mathrm{s1}$ and $R_\mathrm{s2}$. Two schemes of surface current controllability are considered: full control \mbox{$\varOmega_\mathrm{A} = \ell\times\ell/2$} (red curves) and partial control \mbox{$\varOmega_\mathrm{A} = \ell_1 \times \ell/2$} (black curves). Resistivity of the larger area is constant while the resistivity of smaller area is $0\,\Ohm$ (PEC) or $1\,\Ohm$. The selected Pareto optimal current densities are depicted only for relevant part of the plate, highlighted by the dashed yellow region.}
\label{fig:inhomogen}%
\end{figure}

The last example in this paper studies simple rectangular plate, divided into two regions of different surface resistivity, $R_\mathrm{s1}$, $R_\mathrm{s2}$, and considering two different schemes of controllability -- in one only a small area is controllable while in the other the entire plate is controllable. The results are depicted in Fig.~\ref{fig:inhomogen} and it is immediately seen that the Pareto fronts sweep a broad range of dissipation factor and radiation Q-factor values. Notice that the ends of the Pareto fronts corresponding with minimum Q-factor are already known: the optimal current for \mbox{$\varOmega_\mathrm{A}=\ell\times\ell/2$}, \mbox{$R_\mathrm{s1}=1\,\Ohm$} is depicted in \cite{Capek+Jelinek2016} and the optimal current for partial control and PEC is depicted in \cite{Capek+etal2017b}.

Naturally, many permutations of the controllable region problem exist and general conclusions cannot be drawn from a single example.  This formulation is adaptable to the study of bounds and trade-offs of arbitrary antenna / ground configurations and is powerful in determining the specific behavior of optimal currents in complex design settings.  Additional investigation is enabled by considering inhomogeneous loss properties through the use of~\eqref{eq:losses-nonuniform}.  In this way, it is possible to study the impacts of the performance bounds achievable in the region~$\varOmega_\mathrm{A}$ in the presence of an arbitrarily lossy parasitic region~$\varOmega_\mathrm{G}$.  This analysis is of practical importance for informing the specification and design of antennas placed in operating environments where varying loading or parasitic effects may be encountered, \eg{}, in-body or embedded antennas~\cite{Merli+etal2011}.

\section{Conclusion}
\label{sec:conclusions}

Using tools from convex and multi-objective optimization, we study the trade-offs between antenna efficiency and radiation \mbox{Q-factor}. Even with this focused scope, there exists a plethora of constraints (resonance, restricted controllable regions) to explore with minimal changes in the formulation.  Manipulating the choice of operators and different constraints yields many more tractable problems than covered here.  Thus a general conclusion of this work is that, independent of specific numerical outcomes, the proposed formulation of multi-objective current optimization problems enables the study and quantification of many antenna problems and will seed significant further work in this area.
 
More specific conclusions and physical insights are drawn from the numerical examples.  Foremost is the enormous cost of self-resonance on efficiency, \ie{}, the maximum efficiency attainable by a self-resonant current is generally much lower than that achievable by a non-resonant current.  Additionally, though technically continuous, the Pareto front in this problem is clustered into two distinct regions prioritizing either efficiency or \mbox{Q-factor}.  Examination of these clusters shows the rapid transition between a nearly constant current (non-resonant, high efficiency, extremely high Q-factor) and a multi-mode solution (self-resonant, low-efficiency, low \mbox{Q-factor}).  Moving slightly away from the extreme case of the optimal efficiency constant current, we find a smoother single-mode solution which has a \mbox{Q-factor} only slightly higher than that of the multi-mode solution while having significantly improved efficiency when lossless tuning is assumed.  The numerical comparisons between these two classes of solutions vary between examples, but the structure of the results stays the same.  

Given these results we conclude by answering the question: \textit{By looking for the most efficient antenna possible, do we sacrifice Q-factor?}  Yes, but numerical results suggest that the relative tradeoff depends highly on the specific problem under consideration.  For the fully controllable regions studied here, the sacrifice is small, with increases to Q-factor on the order of $10\%$ to $50\%$ allowing for very large increases in efficiency. However for small controllable antenna regions near larger parasitic structures the tradeoff cost between these parameters may be much higher.

\appendices

\section{Degeneracies}
\label{app:degeneracies}
Degenerate eigenvalues require special care when constructing the Pareto front for the tuned problem~\eqref{eq:opt_effQ_tuned}. Many degenerate eigenvalues are associated with symmetries in the geometry of the problem \cite{Schab+Bernhard2016}. Inversion symmetry $\rv\to-\rv$ are inherent in canonical geometries such as rectangles and spheroids, as well as simple antenna geometries like loops and dipoles.  For these cases the symmetries are known and can be used to turn the computational problems with degeneracies to advantages. The inversion symmetry decomposes the solution set into two orthogonal subspaces consisting of symmetric (even) and anti-symmetric current densities. This reduces the size of the matrices and hence computational complexity in the generalized eigenvalue problems. The decomposition also separates the solutions into TM (even) and TE (odd) parts in the limit of electrically small structures.

\ifCLASSOPTIONcaptionsoff
\newpage
\fi

\bibliographystyle{IEEEtran}
\bibliography{total}


\begin{biography}[{\includegraphics[width=1in,height=1.25in,clip,keepaspectratio]{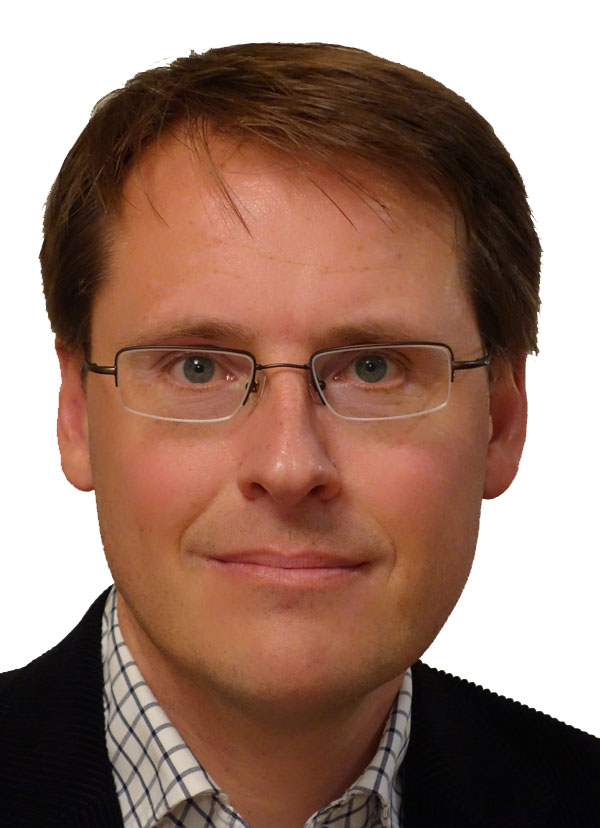}}]{Mats Gustafsson}
received the M.Sc. degree in Engineering Physics 1994, the Ph.D. degree in Electromagnetic
Theory 2000, was appointed Docent 2005, and Professor of Electromagnetic Theory 2011, all
from Lund University, Sweden.

He co-founded the company Phase holographic imaging AB in 2004. His research interests are in scattering and antenna theory and inverse scattering and imaging. He has written over 90 peer reviewed journal papers and over 100 conference papers. Prof. Gustafsson received the IEEE Schelkunoff Transactions Prize Paper Award 2010 and Best Paper Awards at EuCAP 2007 and 2013. He served as an IEEE AP-S Distinguished Lecturer for 2013-15.
\end{biography}

\begin{biography}[{\includegraphics[width=1in,height=1.25in,clip,keepaspectratio]{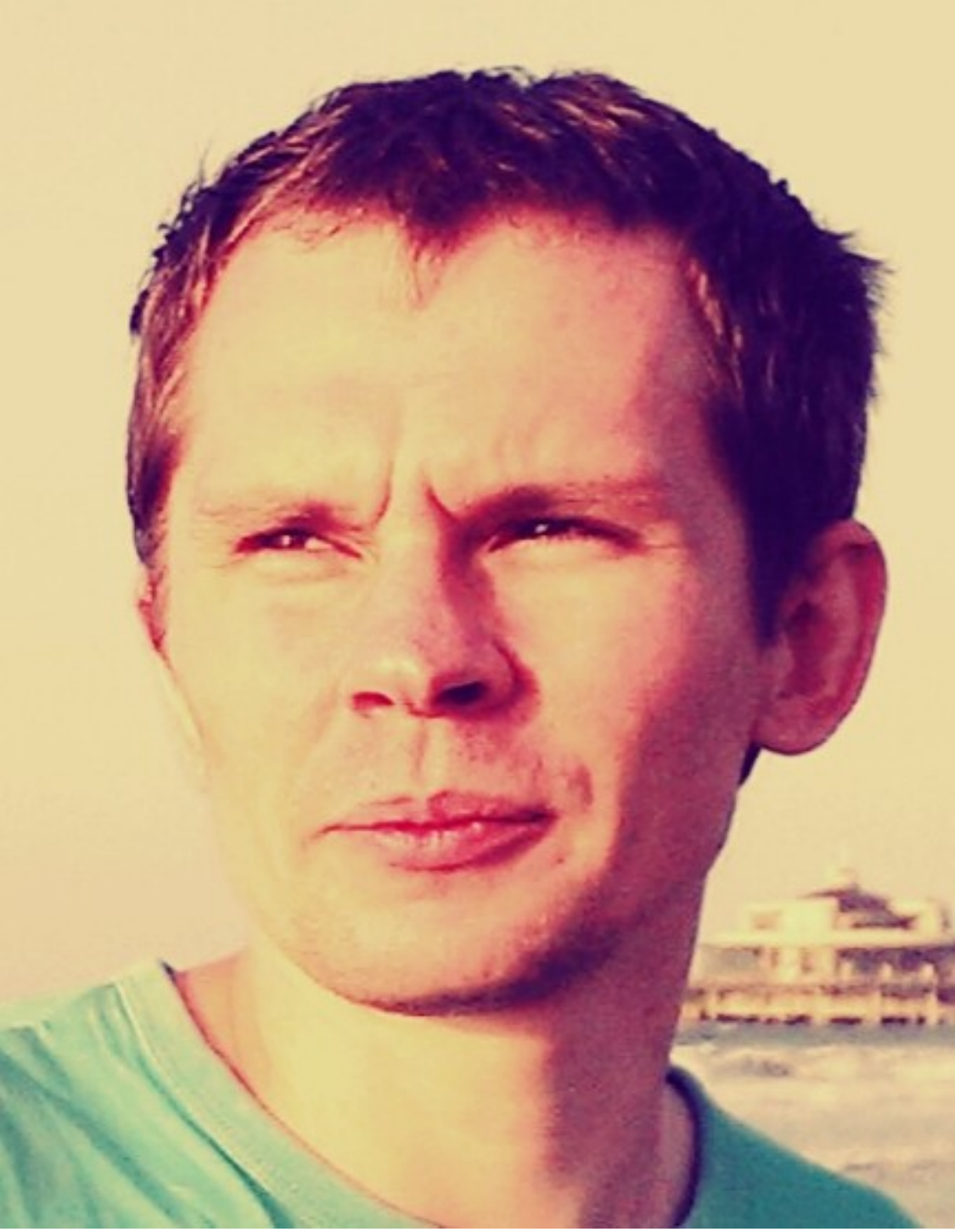}}]{Miloslav Capek}
(M'14, SM'17) received his M.Sc. degree in Electrical Engineering and Ph.D. degree from the Czech Technical University, Czech Republic, in 2009 and 2014, respectively. In 2017 he was appointed Associate Professor at the Department of Electromagnetic Field at the CTU in Prague.
	
He leads the development of the AToM (Antenna Toolbox for Matlab) package. His research interests are in the area of electromagnetic theory, electrically small antennas, numerical techniques, fractal geometry and optimization. He authored or co-authored over 70 journal and conference papers.

Dr. Capek is member of Radioengineering Society, regional delegate of EurAAP, and Associate Editor of Radioengineering.
\end{biography}

\begin{biography}[{\includegraphics[width=1in,height=1.25in,clip,keepaspectratio]{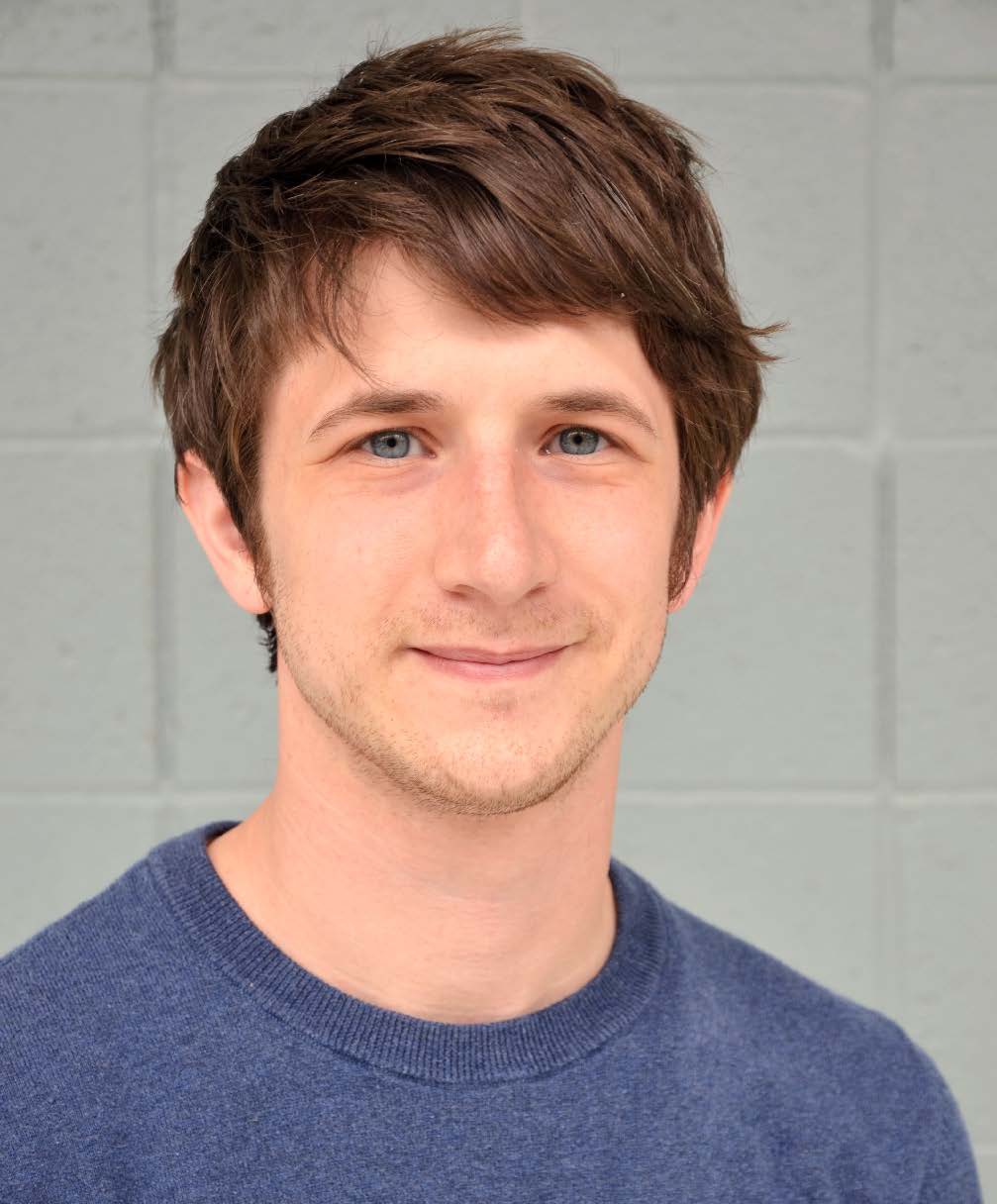}}]{Kurt Schab}
(S'09, M'16) received the B.S. degree in electrical engineering and physics from Portland State University, Portland, OR, USA, in 2011, and the M.S. and Ph.D. degrees in electrical engineering from
the University of Illinois at Urbana-Champaign, Champaign, IL, USA, in 2013 and 2016, respectively.  Currently, he is a postdoctoral research fellow at North Carolina State University.

His research interests include electromagnetic scattering theory, optimized antenna design, and numerical methods in electromagnetics.
\end{biography}

\end{document}